\documentclass[fleqn,usenatbib]{mnras}

\usepackage{newtxtext,newtxmath}
\usepackage[T1]{fontenc}

\DeclareRobustCommand{\VAN}[3]{#2}
\let\VANthebibliography\thebibliography
\def\thebibliography{\DeclareRobustCommand{\VAN}[3]{##3}\VANthebibliography}

\usepackage{graphicx}           
\usepackage{amsmath}            
\usepackage{mathrsfs,bbm}       
\usepackage[caption=false]{subfig}
\usepackage{xcolor}             
\usepackage{hyperref}
\usepackage{tabularx}

\usepackage{lineno}

\newcommand{\vect}{\ensuremath{\textbf{r}}}

\title[Inertial sub-range assessment at GMT and TMT sites]{Inertial sub-range detection and characterization for ground-layer
turbulence at new extremely large telescope sites in Chile}

\author[M. Sep\'ulveda \& D. G. P\'erez]{
Marco Sep\'ulveda$^{1}$
and Dar\'io G. P\'erez$^{1}$\thanks{E-mail: dario.perez@pucv.cl}
\\
$^{1}$Instituto de F\'isica, Pontificia Universidad Cat\'olica de Valpara\'iso (PUCV),
234-0025 Valpara\'iso, Chile
}

\date{Accepted XXX. Received YYY; in original form ZZZ}

\pubyear{2026}

\begin{document}
\label{firstpage}
\pagerange{\pageref{firstpage}--\pageref{lastpage}}
\maketitle

\begin{abstract}
The atmospheric boundary layer (ABL) presents turbulence regimes that diverge from the classical Kolmogorov 1941 (K41) power-law spectrum widely assumed in optical atmospheric profiling, yet turbulence characterization for astronomical applications remains predominantly based on K41 assumptions. Using high-frequency sonic anemometer data from site-testing campaigns at the Giant Magellan Telescope (GMT) and Thirty Meter Telescope (TMT) sites in Chile, we apply detrended fluctuation analysis (DFA)---a technique from fractal time-series analysis---to identify the inertial sub-range and evaluate statistical isotropy in the ABL, comparing our approach with the Fourier-based algorithm for robust identification of the inertial subrange (ARIIS). Strict Kolmogorov conditions---simultaneous isotropy and $k^{-5/3}$ scaling across all fields---are satisfied in fewer than 7 per cent of observations. Temperature fields systematically decouple from velocity components under stable stratification, exhibiting Bolgiano-Obukhov scaling ($k^{-7/5}$) during nighttime hours, indicating active rather than passive scalar behaviour. DFA-detected regime transitions correlate with characteristic turbulence scales, providing objective, model-free boundaries for the inertial subrange. Comparative $C_n^2$ estimation reveals that structure function methods assuming K41 scaling diverge significantly from flux-based parameterizations under stable conditions. The rarity of true Kolmogorov conditions at these sites necessitates stratification-aware methodologies for accurate optical turbulence characterization at next-generation extremely large telescope facilities.
\end{abstract}

\begin{keywords}
atmospheric effects -- site testing -- turbulence -- methods: data analysis --
instrumentation: adaptive optics -- telescopes
\end{keywords}


\section{Introduction}
\label{sec:intro}

One of the most important concepts in turbulence theory is the existence of an inertial subrange. Kolmogorov's celebrated 1941 theory (K41) indicates that whenever the fluctuations of the velocity field, $\delta\mathbf{v}=(u',v',w')$, are statistically isotropic and homogeneous, the structure function follows
\begin{equation}
  S_2(\vect) \equiv \left\langle \left\|\delta\mathbf{v}(\mathbf{r}_0+\mathbf{r})-\delta\mathbf{v}(\mathbf{r}_0)\right\|^2 \right\rangle = C_v^2r^{2/3},
  \label{eq:strucfun}
\end{equation}
where $C_v^2$ is the structure constant representing the velocity fluctuation strength. The variables $u'$ and $v'$ are the streamwise and spanwise (cross-wind) velocity components, respectively, while $w'$ is the vertical wind component perpendicular to the ground. The corresponding one-dimensional power spectrum of the velocity field follows $\Phi_v(\kappa) \propto \kappa^{-\beta}$ with $\beta = 5/3$. Years later, \cite{paper:obukhov1959} obtained identical properties for passive scalar fields, including temperature, and subsequently \cite{book:tatarskii1961} applied these findings to analyze refractive index fluctuations for understanding the effects of turbulence on electromagnetic propagation. Based on these principles, \cite{paper:fried1966a} introduced the coherence diameter $r_0$---the \textit{Fried parameter}---quantifying the effect of atmospheric turbulence on seeing, which has since become the standard metric for site evaluation, telescope design, and adaptive optics performance estimation.

Adaptive Optics (AO) has revolutionized ground-based astronomy by providing real-time compensation for atmospheric turbulence. However, its effectiveness on new generations of giant segmented telescopes depends critically on our understanding of turbulence statistics in the atmospheric boundary layer (ABL). The classical approach assumes that turbulence follows Kolmogorov statistics uniformly throughout the atmosphere, but this assumption may not hold in the complex dynamics of the ABL.
 
How often do K41 conditions hold at the surface of these sites, and how large an error follows when they are assumed where they fail? Answering either question requires determining from the data alone the frequency range over which K41 holds. Aero-thermal models supporting enclosure design take Kolmogorov statistics as given, and estimates of the refractive-index structure constant $C_n^2$ from sonic anemometry \citep{paper:travouillon2015} rest on an unverified inertial subrange. Evidence of departures is scattered across instruments and sites. In particular, sonic-anemometer temperature spectra at Paranal return structure-function exponents well below the $5/3$ value \citep{paper:lombardi2010}, and current profiler intercomparisons still name non-Kolmogorov surface-layer turbulence as a source of disagreement between monitors \citep{paper:berdja2011,paper:griffiths2024}.

Although visual inspection can identify the inertial subrange, it offers no objective or reproducible criterion. A recent solution is the \textit{algorithm for robust identification of the inertial subrange} (ARIIS) proposed by \cite{paper:ortiz-suslow2020}, which conditions the identification of the subrange on statistical isotropy rather than on the $\kappa^{-5/3}$ slope alone. 

Otherwise, Kolmogorov spectra (either for velocity or temperature fields) are univocally associated with self-similar random fields \citep{book:embrechts2002, paper:perez2008}; consequently, the (temporal) power spectrum exponent relates to the \textit{generalized Hurst exponent} following
\begin{equation}
\beta = 2H+1
\label{eq:betahurst}
\end{equation} 
---even for non-stationary cases~\citep{paper:movahed2006, paper:kiyono2015}. Assuming Gaussianity and validity of Taylor's frozen hypothesis, one can model the structure function as a fractional Brownian motion (fBm); since refractive-index fluctuations inherit their statistics from the temperature field, this yields the temporal structure function
\begin{equation}
D_n(\tau) \equiv \left\langle \left[n(\tau)-n(0)\right]^2 \right\rangle \propto \tau^{2H},
\label{eq:increments}
\end{equation}
where $0<H<1$, recovering the Obukhov-Kolmogorov result $\tau^{2/3}$ when $H=1/3$. Among the several techniques available to extract the Hurst exponent, \textit{detrended fluctuation analysis} (DFA), introduced by \cite{paper:peng1994}, stands out for recovering $H$ from non-stationary records without assuming a spectral model beforehand. \cite{paper:funes2016} applied it to laser beam wandering time series and demonstrated experimental evidence of controlled anisotropic turbulence; however, DFA applications for anisotropic analysis under ABL conditions and on meteorological data remain unexplored.

As ARIIS returns, by construction, the exponents of a single identified subrange per spectrum \citep{paper:ortiz-suslow2020}, any multiscale structure of the record remains hidden; DFA acts directly on the anemometric time series and resolves the scales at which the scaling changes \citep{paper:kantelhardt2001, paper:kavasseri2004}. Over complex terrain the distinction carries weight, because spectral isotropy \citep{paper:kurien2001}, the transitions between scaling regimes \citep{paper:cheng2020}, and the coupling between velocity and temperature \citep{paper:wyngaard-cote1971} all respond to stratification, and each has been examined on its own even though the passive-scalar assumption that carries temperature statistics into $C_n^2$ requires the three to hold at once \citep{paper:wyngaard1971, paper:obukhov1959}. That assumption survives only when mechanical forcing dominates buoyancy in the surface layer, a balance captured by \textit{Monin-Obukhov similarity theory} (MOST) \citep{paper:hartogensis2005, paper:li2012, paper:stiperski2023}.

In this work we analyse anemometric records from the GMT and TMT sites, where the next generation of extremely large telescopes is under construction, answering the call for in-situ characterisation of surface-layer turbulence at these sites. DFA is applied to boundary-layer turbulence for the first time outside controlled conditions, while ARIIS receives its first application at an astronomical site, and each is read against the other over the same records to establish what it resolves. Section~\ref{sec:methods} describes the campaigns, the two methods and the processing they share; Sec.~\ref{sec:results} reports how often K41 holds at each site, where the scaling changes and how the choice of assumption propagates into $C_n^2$. Section~\ref{sec:discussion} interprets these results physically and states the limitations of the approach, and Sec.~\ref{sec:conclusions} returns to the questions posed here.

\section{Data and methods}
\label{sec:methods}

\begin{figure}
  \centering
  \includegraphics[width=\hsize]{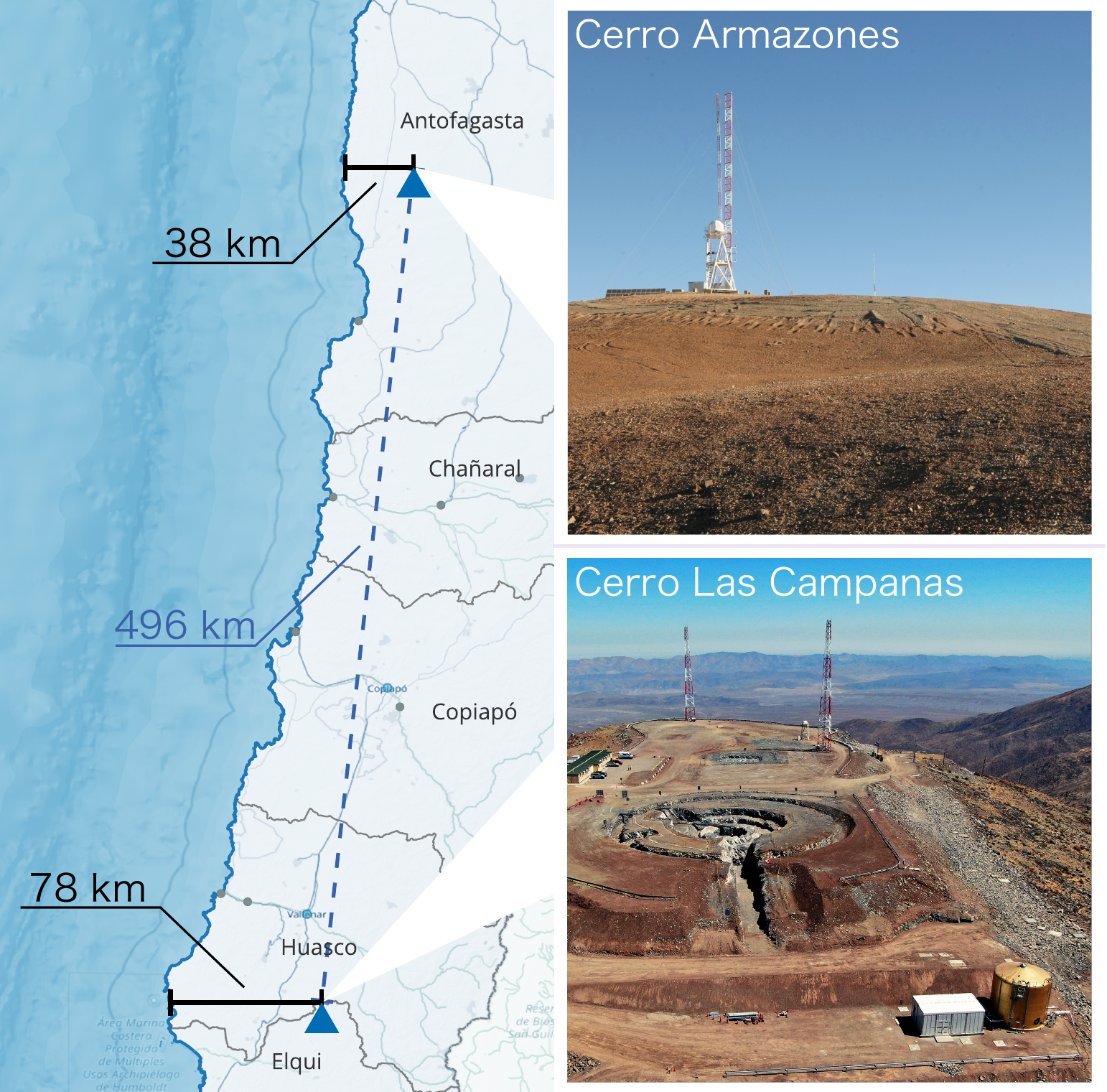}
  \caption{Location of the TMT and GMT anemometry campaigns within Chilean territory. The inter-site separation is approximately 496 km.}
  \label{fig:map}
\end{figure}

\subsection{Sites, instrumentation and datasets}
\label{sec:Dataset}

The TMT dataset comprises five years of atmospheric measurements collected on five mountains across North and South America using an unprecedented array of instrumentation for the site selection and evaluation of what was the planned Thirty Meter Telescope and the currently developing Extremely Large Telescope (ELT)---for additional details on this dataset, see \citet{paper:schock2009} and \citet{paper:travouillon2015}. In this work, we specifically analyze wind velocity components $u$, $v$, $w$, and temperature $T$ data recorded by a CSAT3 sonic anemometer (Campbell Scientific) with a $60$ Hz sampling rate, positioned on Cerro Armazones ($24^\circ 35'22.00''$S $70^\circ 11'31.00''$W, ELT placement) at $3064$ m above sea level and $28$ m above ground in Antofagasta, Chile, during December 2008. For the GMT dataset, we analyzed the same variables collected by a Young 81000 model anemometer (R.M. Young Company) operating at a $30$ Hz sampling rate. It was located at the construction site of the Giant Magellan Telescope (GMT) in Las Campanas Observatory, Chile ($29^\circ02'54.0''$S $70^\circ 41'01.0''$W). The anemometer was installed at a height of $37$ m above ground and $2514$ m above sea level, with data acquisition occurring continuously from December 1st, 2022, through January 31st, 2023. The wind components allow us to analyze the state of statistical isotropy (anisotropy), while the temperature gives indirect access to the fluctuations of the refractive index. Cerro Armazones is situated approximately $38$ km inland from the Pacific Ocean, while Las Campanas Observatory is about $78$ km from the coast, and the two sites are separated by approximately $496$ km (see Fig.~\ref{fig:map}).

The output datasets from both sonic anemometers were initially recorded in geographic coordinates and subsequently rotated to streamwise coordinates following the methodology described in \citet{url:ncar}. Neither campaign provided specific information on the anemometer tilt relative to the local surface normal; therefore, we  assumed negligible instrument tilt and applied only the rotations required to transform the data to a streamwise reference system. Scaling exponents $H$ were computed from consecutive, non-overlapping $30$-minute time series, yielding $48$ Hurst exponents per variable and per day. For the TMT campaign, data acquisition began at 0930 UTC. For GMT, acquisition began at 1900 UTC on December 01, 2022, and at 0001 UTC on January 01, 2023. All times in this work are UTC, three hours ahead of local time during the campaigns. Each site contributes 21 days to the statistics of Sec.~\ref{sec:results}.

To ensure the applicability of Taylor's hypothesis, $r=\overline{U}\tau$, where $\overline{U}$ is the mean streamwise wind speed, we imposed $\langle u'^2\rangle/\overline{U}^2<0.5$ over $1$-minute intervals \citep{paper:lin1953}. Structure constants were estimated from the structure function over the intervals satisfying this criterion. The scales at which the DFA scaling changes (\textit{crossovers}, defined below) were instead obtained from $30$-minute averages, with each block represented by the mean of its qualifying $1$-minute estimates. We classified a record as Kolmogorov when all Hurst exponents satisfied $0.28\leq H_j\leq0.38$ for $j=u,v,w,T$, corresponding to a $15\%$ band around the theoretical value $H=1/3$.

\subsection{The DFA fluctuation function and crossover detection}
\label{sec:DFA}

Unlike spectral analysis, detrended fluctuation analysis (DFA) is applicable to both stationary and non-stationary time series, which makes it a more appropriate technique when uncertainty exists about the stationarity of the data. It recovers the scaling of a record carrying trends of unknown origin and shape. Given a time series $x_k = \{x_1,x_2,\dots,x_N\}$, the Hurst exponent is recovered in four steps \citep{paper:kantelhardt2001, paper:perez2024}:
\begin{enumerate}
    \item determine the profile $Y(i)=\sum_{k=1}^{i}x_k-\langle x \rangle$, the cumulative sum of the demeaned series; this integration step is omitted when the series is already cumulative;
    \item divide $Y(i)$ into $N_s \equiv [N/s]$ non-overlapping windows of equal length $s$, with $N$ the number of samples; repeating the partition from the opposite end retains the remainder and gives $2N_s$ windows;
    
    \item for each window $\tilde{\nu} = 1, \dots, 2N_s$, fit a second-order polynomial $p_{\tilde{\nu}}$ to the profile by ordinary least squares (hence DFA-2), with $p_{\tilde{\nu}}(j)$ its value at the $j$-th sample of the window, $j = 1, \dots, s$. The \emph{detrended variance} of the window, the mean squared residual of this fit,
\begin{equation*}
F_s^2(\tilde{\nu}) = \frac{1}{s} \sum_{j=1}^{s} \left[ Y(\tilde{\nu}s + j - s) - p_{\tilde{\nu}}(j) \right]^2 ,
\end{equation*}
measures the departure of the profile from its local trend in that window (the windows taken from the opposite end are indexed from the end of the record);

    \item finally, average $F_s^2(\tilde{\nu})$ over the $2N_s$ windows and take the square root to obtain the DFA fluctuation function
\begin{equation*}
F_2(s) = \left[ \frac{1}{2N_s} \sum_{\tilde{\nu}=1}^{2N_s} F_s^2(\tilde{\nu}) \right]^{1/2} ,
\end{equation*}
the root-mean-square fluctuation of the whole record at scale $s$.
\end{enumerate}

If the time series comes from a self-similar process with exponent $H$, the fluctuation function will scale as
\begin{equation}
    F_2(s)\sim s^{H}
    \label{eq:fluctuations}
\end{equation}
for $s$ large enough---a $\log$-$\log$ linear fit usually reveals this trend. In practice, $H$ is obtained as the slope of an ordinary least-squares fit of $\log_2 F_2(s)$ against $\log_2 s$ over a defined scale window.

The structure function and $F_2(s)$  are related,  with the latter combining increments over separations shorter than $s$ through weights determined by the detrending procedure \citep{paper:lovsletten2017}. For self-similar processes both return the same exponent, so the DFA estimate of $H$ carries into the structure-function scaling of equation~\eqref{eq:increments} and the spectral exponent of equation~\eqref{eq:betahurst}. In general, the structure function at a given separation still retains contributions from larger scales \citep{paper:huangy2010}, whereas detrending confines $F_2(s)$ to its window, so a change of slope with $s$ reflects the flow at that scale. 

\begin{figure}
   \centering
   \includegraphics[width=\hsize]{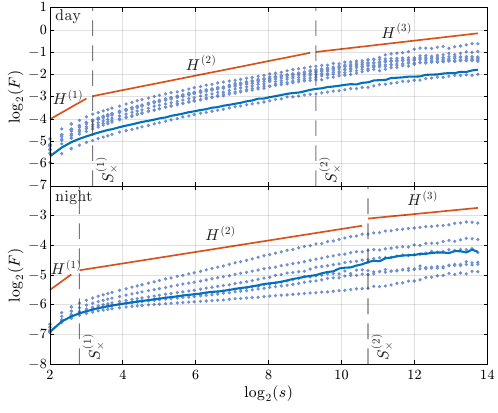}
\caption{Crossover estimation example for GMT temperature records of January 14th, 2023, by day (top, 09:30--14:30 local time) and at night (bottom, 04:30--06:30 local time). The heavy line is the fitted 30-minute record, with $H^{(1)}, H^{(2)}, H^{(3)}$ equal to $0.93, 0.33, 0.19$ (top) and $0.89, 0.19, 0.12$ (bottom); the dots are the other records; $S_{\times}^{(1)}$ and $S_{\times}^{(2)}$ follow from the segmented regression of \citet{paper:ge2013}.}
      \label{Fig:crossTime}
\end{figure}

Frequently, real data exhibit different scaling regimes with increasing $s$, separated by crossover time scales $S_{\times}$ \citep{paper:kavasseri2004, paper:liu2022, paper:movahed2006}, and such multi-regime behaviour is commonplace in turbulence assessment as it reflects the interplay between different turbulence-generating mechanisms. In the surface layer we anticipate two such scales, one set by external forcing and the other by viscous dissipation, which bracket the candidate inertial subrange between them. While these crossover points have often been estimated by ``eyeballing'', efforts have been made to enhance accuracy in their determination~\citep{paper:ge2013, paper:mesquita2021, paper:killick2012}. Following \citet{paper:ge2013}, we employ a segmented regression model where optimal crossover points minimize the total mean squared error across three linear segments fitted to all possible two-breakpoint configurations, enabling objective identification of regime changes in scaling behaviour. The corresponding frequencies are $f_L^\text{DFA} = f_s/S_{\times}^{(2)}$ and $f_H^\text{DFA} = f_s/S_{\times}^{(1)}$, where $f_s$ is the sampling frequency and $f_L^\text{DFA}$, $f_H^\text{DFA}$ represent the low and high crossover frequencies.  These two crossover scales define three scaling exponents, $H^{(1)}$ for the small-scale regime ($s < S_{\times}^{(1)}$), $H^{(2)}$ for the candidate inertial subrange ($S_{\times}^{(1)}<s<S_{\times}^{(2)}$), and $H^{(3)}$ for the large-scale regime ($s>S_{\times}^{(2)}$), as illustrated in Fig.~\ref{Fig:crossTime}. The interval bounded by these two frequencies, $\Delta f^{\text{DFA}} = [f_L^\text{DFA}, f_H^\text{DFA}]$, is the DFA bandwidth where we expect to observe the inertial subrange.  At the smallest windows the fluctuation function departs from linear scaling, since few samples enter each polynomial fit \citep{paper:kantelhardt2001, paper:carpena2021}. The departure places a floor on the detectable high crossover, so $S_{\times}^{(1)}$  can be set by the sampling rate rather than by the flow.

Whenever isotropy and homogeneity are observed in a given spatial (temporal) range, all the velocity components should produce $H=1/3$, equation~\eqref{eq:fluctuations}---i.e., $\tau^{2/3}$ from equation~\eqref{eq:increments}---, indicating the presence of Kolmogorov turbulence. The intensity of this turbulence can be characterized by the structure constant $C^2_v$. Likewise, a predominant Obukhov-Kolmogorov (OK) behavior for the turbulent fluctuations of both temperature and refractive index fields will produce the same Hurst exponent, with the strength of the latter fluctuations ultimately characterized by $C^2_n$.

Conversely, persistent deviations from $H=1/3$---for temperature, refractive index, or any proxy of them---identify the non-Obukhov-Kolmogorov (NOK) regimes, for which isotropic spectral models have been introduced,
\begin{equation}
\Phi_n(\kappa)=A(\alpha)\tilde{C}_n^2(\alpha)\kappa^{-\alpha+2},  \text{ with } 3<\alpha < 4, \label{eq:nonk}
\end{equation}
where $A(\alpha)$ is a dimensionless constant, and $\tilde{C}_n^2(\alpha)$ is a \textit{generalized structure constant} with units m${}^{-\alpha+3}$---here $\alpha = \beta +2$ and thus $\alpha = 2H+3$. This spectrum has often been used to feed wavefront propagation equations to model relevant quantities of interest, such as scintillation and wandering. Nevertheless, a phenomenological connection between the generalized structure constant and the actual structure constant, $C^2_n$, is nonexistent.

\subsection{Inertial range characterization through ARIIS}
\label{sec:ARIIS}

Originally developed for the marine boundary layer, the algorithm for robust identification of the inertial subrange (ARIIS) is applied here to a terrestrial one for the first time. Rather than fitting a slope, it tests whether the streamwise and transverse velocity spectra stand in the ratio that isotropy requires. Upon analyzing data from a sonic anemometer, ARIIS outputs the power exponents for the three velocity components, and the isotropy coefficient $\mathcal{J}$, introduced by \citet{paper:jimenez1993}, given by
\begin{equation}
    \mathcal{J} = \frac{E_{11}(\kappa_1)-\kappa_{1}\frac{\partial E_{11}(\kappa_{1})}{\partial \kappa_1}}{2E_{22}(\kappa_1)},
    \label{eq:isotropy}
\end{equation}
where $E_{11}(\kappa_1)$ and $E_{22}(\kappa_1)$ are longitudinal and transverse one-dimensional spectra, respectively.

In this context, isotropy in atmospheric turbulence occurs when $\mathcal{J}(\Delta f) \rightarrow 1$. Therefore, for ARIIS, a subrange within any spectrum exhibiting $\kappa^{-5/3}$ scaling alone may mischaracterize the inertial subrange, leading to incorrect identification of turbulence characteristics within the spectral bandwidth. The condition of isotropy must be satisfied because of Kolmogorov's original argument that the turbulent motions over this subrange are independent from the inherently anisotropic large-scale mean motions. Additionally, the transformation from temporal frequency to spatial frequency requires invoking Taylor's frozen turbulence hypothesis. However, using temporal frequency instead of spatial frequency preserves the fundamental characteristics of the Kolmogorov spectrum slope, and thus the inertial subrange can be defined by analyzing the temporal variable \citep{paper:ortiz-suslow2021}.

Through equation~\eqref{eq:isotropy}, ARIIS identifies the isotropy band $\Delta f^\text{ARIIS}=[f_{L}^{\text{ARIIS}}, f_{H}^{\text{ARIIS}}]$ under the criterion $0.95 \leq \mathcal{J} \leq 1.05$ and fits spectral exponents $m_{uu}$, $m_{vv}$, $m_{ww}$, and $m_{TT}$ for the streamwise, transverse, vertical, and temperature fields, converted to their Hurst counterparts $H_u^{\mathrm{ARIIS}}$, $H_v^{\mathrm{ARIIS}}$, $H_w^{\mathrm{ARIIS}}$, and $H_T^{\mathrm{ARIIS}}$ via equation~\eqref{eq:betahurst}. In the implementation used here the band starts no lower than $U/z$, its upper edge is fixed, only its inner 80 per cent is retained, and the spectra are corrected for sonic path averaging, which the DFA time series are not. DFA offers a weaker counterpart: a record is \textit{isotropic in the DFA sense} when $H_j = 1/3 \pm 15\%$ holds simultaneously for all velocity components. A common exponent is necessary for K41 isotropy, equation~\eqref{eq:strucfun}, but not sufficient, since only $\mathcal{J}$ tests the amplitude ratio between the components.

\subsection{Stability regimes and characteristic turbulence scales}
\label{sec:MOST}

Whether temperature behaves as a passive tracer or actively contributes to turbulence generation is decided by the competition between mechanical and buoyant production of turbulence, which \textit{Monin-Obukhov similarity theory} (MOST) quantifies. Temperature fluctuations are linked to buoyancy and wind shear through the sensible heat flux $\overline{w'\theta'}$ and friction velocity $u_*=(\overline{u'w'}^2+\overline{v'w'}^2)^{1/4}$~\citep{paper:pierzyna2024}, where $\theta$ represents the potential temperature. The analysis centers on the dimensionless stability parameter $\zeta = z/L$, where $z$ represents the height above ground and
\begin{equation}
L = -\frac{u_*^3\overline{T}}{\kappa g \overline{w'\theta'}}
\label{eq:Obukhov_length}
\end{equation}
defines the \textit{Obukhov length scale}. This characteristic length represents the height at which buoyant production of turbulence equals mechanical production. Here, $\kappa \sim 0.4$ is the von Kármán constant, and $g$ denotes gravitational acceleration~\citep{paper:li2012,paper:hartogensis2005}.
The sign and magnitude of $\zeta$ indicate different stability regimes: negative values represent unstable conditions where buoyancy enhances turbulence, positive values indicate \textit{stable stratification conditions} ($\zeta > 0$) where buoyancy suppresses turbulence, and values near zero correspond to \textit{near-neutral conditions} where mechanical turbulence dominates. Each record is classified by its stratification regime and read against the scales that regime imposes.

A stratified boundary layer offers no single outer scale. Buoyancy, shear and the temperature field each impose one, and which of them truncates the inertial subrange indicates which mechanism governs the flow. The Ozmidov scale
\begin{equation}
L_o = 2\pi(\epsilon/N^3)^{1/2},
\label{eq:oz_es}
\end{equation}
with $\epsilon$ the dissipation rate of \textit{turbulent kinetic energy} (TKE) and $N$ the Brunt-V\"ais\"al\"a frequency, marks where stratification suppresses three-dimensional overturning; the Obukhov length of equation~\eqref{eq:Obukhov_length} marks where buoyant production overtakes mechanical production; and the Bolgiano length
\begin{equation}
L_B = \epsilon^{5/4}/[\chi^{3/4}(g\alpha_T)^{3/2}],
\label{eq:bo_es}
\end{equation}
with $\chi$ the dissipation rate of temperature variance and $\alpha_T$ the thermal expansion coefficient, marks where buoyancy begins to inject energy directly into the cascade \citep{paper:skrbek2002, book:wyngaard2010, paper:basu2022}. At the opposite end two scales bound the dissipative range, the Taylor microscale
\begin{equation}
\lambda = \sqrt{15\nu u'^2/\epsilon}
\label{eq:tay_es}
\end{equation}
marking the onset of viscous influence and the Kolmogorov scale $\eta = (\nu^3/\epsilon)^{1/4}$ its completion. Each converts to a frequency through the root-mean-square horizontal velocity $U$, giving $f_o = U/L_o$, $f_\text{O}=U/|L|$, $f_B =U/L_B$, $f_\lambda = U/\lambda$ and $f_\eta = U/\eta$, alongside the buoyancy frequency $f_b = N/(2\pi)$, all computed from the same records.

\subsection{$C_n^2$ estimation}
\label{sec:Cn2}

The refractive-index structure constant follows from the temperature field. At the TMT and GMT sites, water-vapor and pressure fluctuations are negligible, so refractive-index fluctuations track temperature fluctuations alone~\citep{paper:andreas1988,paper:frederickson2000}, and the two structure constants are proportional through $C_n^2 = A^2 C_T^2$, where $A = -10^{-6} P T^{-2} m_1(\lambda_0)$ and $m_1(\lambda_0) = 23.7134 + 6839.397/(130-\lambda_0^{-2}) + 45.473/(38.9-\lambda_0^{-2})$, evaluated at $\lambda_0 = 5~\mu$m; since every estimator carries the same factor $A^2$, this choice sets the absolute level without affecting the comparison between them. This proportionality holds whenever temperature behaves as a passive scalar over an existing inertial subrange, the two conditions the estimators below assume to differing degrees. Every value reported here is directly computed from the raw velocity and temperature records. 

The amplitude of the fluctuation function depends on the detrending procedure and carries no turbulence structure constant, so  $C_n^2$ needs to be extracted from the refractive-index structure function. Assuming classical K41 scaling and passive temperature behavior, $(C_n^2)^\text{SF}$ follows from equation~\eqref{eq:strucfun} applied to the temperature field under the spatio-temporal transformation $r=\overline{U}\tau$ introduced in Sec.~\ref{sec:Dataset}, which $89.56\%$ and $88.98\%$ of $1$-minute intervals satisfy for the GMT and TMT campaigns, respectively. Replacing the fixed  $2/3$ exponent with the value recovered in Sec.~\ref{sec:DFA} extends the same construction to the NOK cases, giving 
\begin{equation}
S_2(r) = \tilde C_n^2(\alpha)\, r^{2H},
\label{eq:genstrucfun}
\end{equation}
which produces a structure constant $(C_n^2)^\text{DFA}$ with dimensions $\text{m}^{-2H}$ which are unphysical~\citep{inprocc:sepulveda2021}. 

Similarity theory offers an independent route. Following \citet{paper:wyngaard1971} as implemented by \citet{paper:pierzyna2024} in their intercomparison of ABL $C_n^2$ parameterizations against scintillometer measurements,  the \textit{flux-based} $(C_n^2)^\text{Flux}$ rests on
\begin{equation}
C_T^2 = T_*^2 z^{-2/3} g(\zeta),
\label{Eq:CT2_Pier}
\end{equation}
where $T_* = -\overline{w'\theta'}/u_*$ is the temperature scale and
\begin{equation}
g(\zeta) =4.9\times\begin{cases} (1-6.1 \zeta)^{-2/3}, & \zeta < 0 \\
(1+2.2 \zeta^{2/3}), & \zeta \geq 0
\end{cases}
\label{Eq:g_func}
\end{equation}
is the stability function accounting for stratification effects.

A second similarity route replaces the fluxes by variances. Following \citet{paper:wyngaard1971} and \citet{paper:corrsin1951}, $(C_n^2)^\text{VB}$ follows from the \textit{variance-based} relationship 
\begin{equation}
C_T^2 = 3.2 \chi \epsilon^{-1/3}. 
\label{eq:Cn2_VB}
\end{equation}
Its parameters are expressed through turbulent kinetic energy scaling as $\epsilon = (2e)^{3/2}/(B_1 L_M)$ and $\chi = (2e)^{1/2}\sigma_\theta^2/(B_2 L_M)$, with empirical constants $B_1 = 24$ and $B_2 = 15$ \citep{paper:mellor1974}, where $e = (\sigma_u^2 + \sigma_v^2 + \sigma_w^2)/2$ is the turbulent kinetic energy. The master length scale $L_M$ combines contributions from surface forcing, turbulence and buoyancy through $L_M^{-1} = l_s^{-1} + l_t^{-1} + l_b^{-1}$. Following \citet{book:olson2019} and \citet{paper:blackadar1962}, the surface forcing scale is $l_s = \kappa z$, the turbulence scale is $l_t = 50$~m under stable conditions and $l_t = 500$~m for unstable stratification, and the buoyancy scale is $l_b = 2\pi U/N$ when $\zeta > 0$ and $l_b = \tau e^{1/2}$ for $\zeta \leq 0$, where $\tau = 0.5z_i/w_*$ combines the planetary boundary layer height $z_i$ with the convective velocity scale $w_* = [(g/\theta)z_i\overline{w'\theta'}]^{1/3}$~\citep{paper:wulfmeyer2016,paper:deardorff1970}. For the present datasets, $w_*$ is instead obtained from the empirical relation $w_*=\sqrt{(\sigma_u^2+\sigma_v^2)/0.7}$~\citep{paper:panofsky1977}. Substitution into the relation above gives $\tau = 0.5\,\theta w_*^2/(g\overline{w'\theta'})$, so the estimate no longer requires $z_i$.

Under stable stratification, scaling arguments applied to the budget equations for TKE and scalar variance \citep{paper:basu2022} give
\begin{equation}
(C_n^2)_\text{max}^\text{Basu}  = \left(A\frac{\partial \bar{\theta}}{\partial z}\right)^2 \left(\frac{c}{\text{Pr}_t}\right)  L_X^{4/3},
\label{Eq:Cn2_Sukanta}
\end{equation}
where $c$ is an empirical constant in the range $[2.8,3.2]$, for which we use $c=3.2$, $\text{Pr}_t$ is the local turbulent Prandtl number, taken as unity, and $L_X \simeq 0.46\,\sigma_\theta\sqrt{\text{Pr}_t}/(\partial \bar{\theta}/\partial z)$ is an integral scale with $\partial \bar{\theta}/\partial z = T_{*}\phi(\zeta)/(\kappa z)$ and $\phi(\zeta) = 1 + 7\zeta$ the similarity function for $\zeta > 0$ \citep{paper:basu2022, paper:andreas1988}.

\section{Results}
\label{sec:results}
\begin{figure}
   \centering
   \includegraphics[width=\hsize]{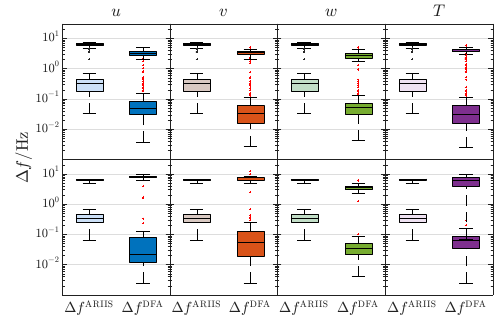}
      \caption{Inertial subrange frequency bounds detected by ARIIS and DFA for velocity components $(u,v,w)$ and temperature in GMT (top row) and TMT (bottom row) campaigns. Statistics include only records isotropic in the DFA sense.}
      \label{fig:temp_inertial}
\end{figure}
The two methods bound the inertial subrange differently. ARIIS produces a uniform frequency band across all quantities, whereas DFA yields quantity-specific bands from the individual fluctuation functions (Fig.~\ref{fig:temp_inertial}). In the GMT dataset the DFA high-frequency boundary is constrained by $S_{\times}^{(1)}$ for the reason given in Sec.~\ref{sec:DFA}. The $60$~Hz TMT data alleviate this, as the median temperature frequency ranges illustrate. GMT gives $f_L = 0.0310 \pm 0.0027$~Hz and $f_H = 3.838 \pm 0.055$~Hz for DFA against $f_L = 0.333 \pm 0.012$~Hz and $f_H = 6.289 \pm 0.055$~Hz for ARIIS, while TMT gives $f_L = 0.0764 \pm 0.0080$~Hz and $f_H = 5.888 \pm 0.334$~Hz against $f_L = 0.352 \pm 0.020$~Hz and $f_H = 6.394 \pm 0.073$~Hz.

\begin{figure}
   \centering
   \includegraphics[width=\hsize]{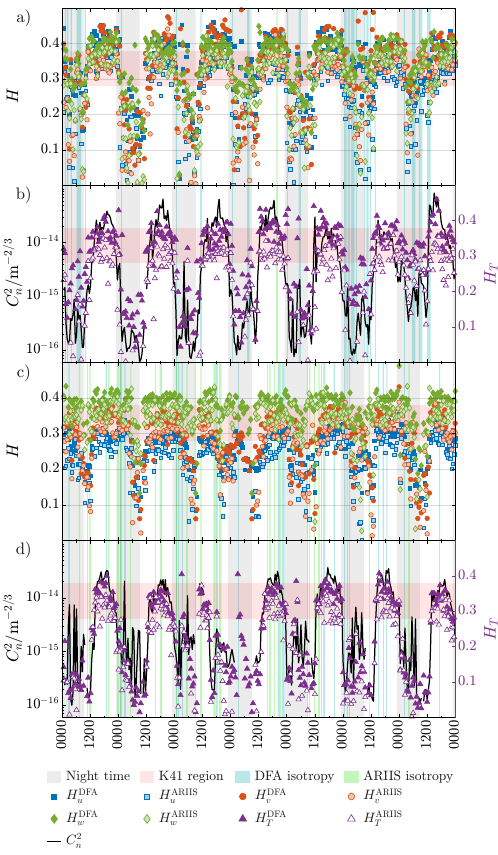}
      \caption{Hurst exponents from DFA (filled symbols) and ARIIS (open symbols) for velocity components $(u,v,w)$ and temperature $T$ during December 1st-7th, UTC. Gray columns indicate nighttime intervals; red bands show the Kolmogorov regime $(H = 1/3 \pm 15\%)$ and light-blue and green columns mark records isotropic in the DFA sense and by ARIIS, respectively. The solid black line in panels (b) and (d) gives $(C_n^2)^\text{SF}$, computed from the temperature records as described in Sec.~\ref{sec:Cn2}. Upper panel (a)-(b): GMT campaign. Lower panel (c)-(d): TMT campaign.}
      \label{fig:DFA_ARIIS}
\end{figure}
\begin{table}
\caption{Pearson correlation coefficients between Hurst exponents obtained by DFA and ARIIS techniques on $u$, $v$, $w$ and $T$ sonic anemometer variables.}
\centering
\begin{tabular}{lcccc}
\hline\hline
Period               &     $u$      &     $v$     &     $w$     &     $T$       \\
\hline
TMT dec 2008    &    $0.46$    &    $0.54$     &    $0.63$     &    $0.63$   \\
GMT dec 2022    &    $0.54$    &    $0.54$     &    $0.67$    &     $0.68$   \\
GMT jan 2023    &    $0.62$     &    $0.47$    &    $0.71$     &    $0.74$   \\
\hline
\end{tabular}
\label{table:TABLA1}
\end{table}
K41 conditions cluster between 0000 and 1200 UTC at both GMT (December 2022) and TMT (December 2008). In Fig.~\ref{fig:DFA_ARIIS} vertical bands mark them, light blue where DFA gives $H_{u,v,w} \sim 1/3$, equation~\eqref{eq:fluctuations}, and green where ARIIS finds isotropy, equation~\eqref{eq:isotropy}. Outside these intervals both methods produce intermediate exponents whose interpretation depends on their values and the associated frequency ranges. The two methods agree in temporal evolution and magnitude, yet the Pearson correlations of Table~\ref{table:TABLA1} reach only $0.46$ to $0.74$. The light-blue bands of Fig.~\ref{fig:DFA_ARIIS} mark the records isotropic in the DFA sense, $10.5$ and $12.4$ per cent of the total at GMT and TMT.
\begin{figure}
        \includegraphics[width=\hsize]{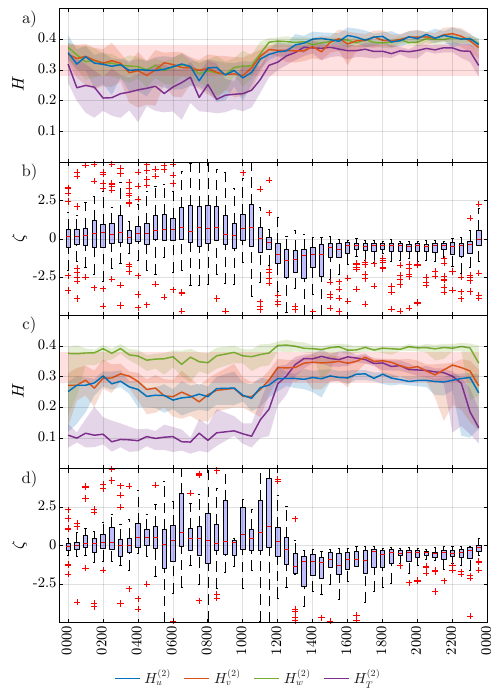}%
\caption{Half-hourly diurnal variation of the DFA Hurst exponent $H^{(2)}$ and the stability parameter $\zeta$ over 21 days per site: (a) median $H^{(2)}$ for velocity components ($u$, $v$, $w$) and temperature $T$ at GMT, with 25th--75th percentile bands; (b) $\zeta$ for GMT, as box plots of each half-hour; (c) median $H^{(2)}$ for TMT; (d) $\zeta$ for TMT. Red bands indicate the Kolmogorov regime ($H = 1/3$ $\pm 15\%$).}
    \label{fig:trend_exponents}
\end{figure}

Figure~\ref{fig:trend_exponents} tracks the candidate inertial-range exponent $H^{(2)}$ against the stability parameter $\zeta = z/L$, with $L$ the Obukhov length of equation~\eqref{eq:Obukhov_length}, through the diurnal cycle. Stable conditions occur mostly at night, and there the temperature exponent leaves the band occupied by the velocity components at both sites, falling to $H_T^{(2)} \simeq 0.2$ or below (Fig.~\ref{Fig:crossTime}, bottom). Under \textit{unstable stratification} ($\zeta < 0$), characteristic of daytime convection, buoyancy enhances turbulence, the coupling is restored and thermal fluctuations approach the K41 exponent. ARIIS returns the same behaviour in Fig.~\ref{fig:DFA_ARIIS}, with $H_T^{\mathrm{ARIIS}}$ shifting from Bolgiano-Obukhov (BO) values under stable conditions toward K41 values under unstable ones and separating from the velocity exponents at night, so the decoupling holds independently of the method used to measure it.
\begin{figure}
   \centering
    \includegraphics[width=.75\hsize]{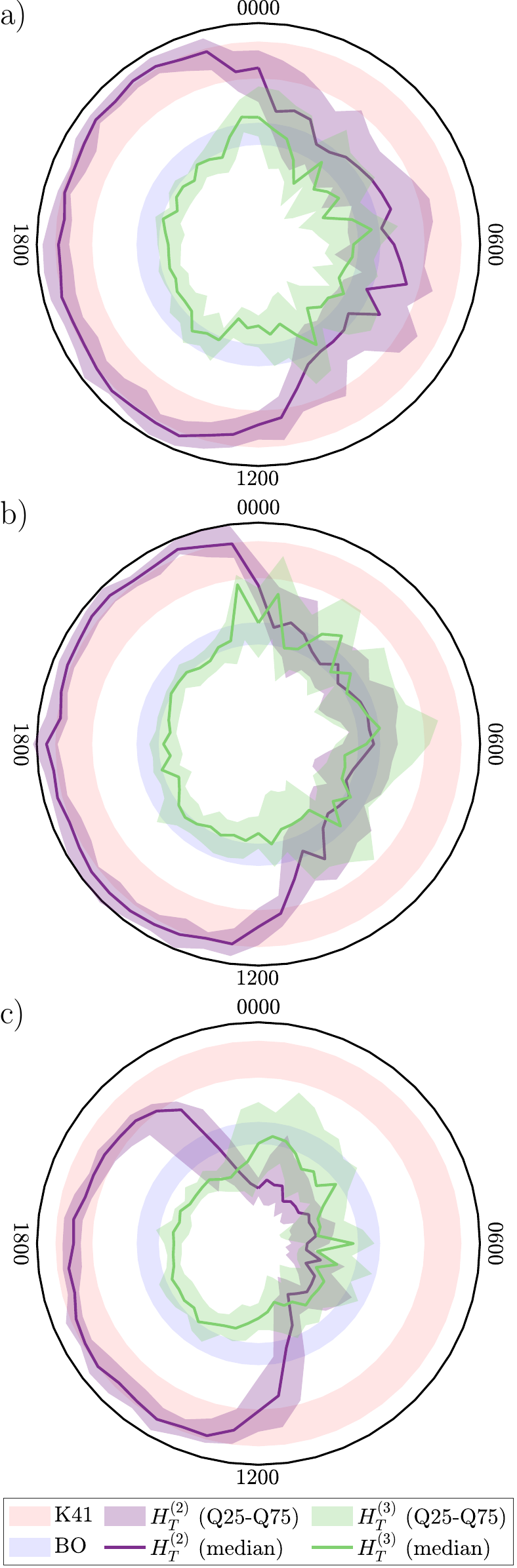}
\caption{Clock plots showing daily patterns of the temperature exponents $H_T^{(2)}$ and $H_T^{(3)}$. Each plot displays median values (solid lines) with interquartile ranges (Q25-Q75, shaded bands) over a 24-hour cycle: (a) GMT December 2022, (b) GMT January 2023, and (c) TMT December 2008.  Red and blue shaded annular regions indicate Kolmogorov (K41, $H=1/3\pm 15\% $) and Bolgiano-Obukhov (BO, $H=0.2\pm 15\%$) scaling regimes, respectively.}
      \label{fig:clock_exponentes}
\end{figure}

The diurnal evolution of scaling exponents reveals distinct multi-scale turbulent regimes in the temperature field, as illustrated in Fig.~\ref{fig:clock_exponentes}. $H_T^{(2)}$ cycles between inertial subrange values during daytime and BO scaling under nocturnal stratified conditions, with BO-compatible values occurring in roughly $60$ per cent of nighttime intervals, while the GMT dataset shows that velocity field exponents lack the corresponding $H = 0.6$ ($k^{-11/5}$) predicted by BO theory. The low-frequency exponent $H_T^{(3)}$ alternates between BO-like and K41-like values at larger scales. The TMT dataset presents an even more intriguing scenario, with temperature Hurst exponents frequently falling below $0.1$, corresponding to power spectra following approximately $k^{-1.2}$.

\begin{figure}
   \centering
    \includegraphics[width=\hsize]{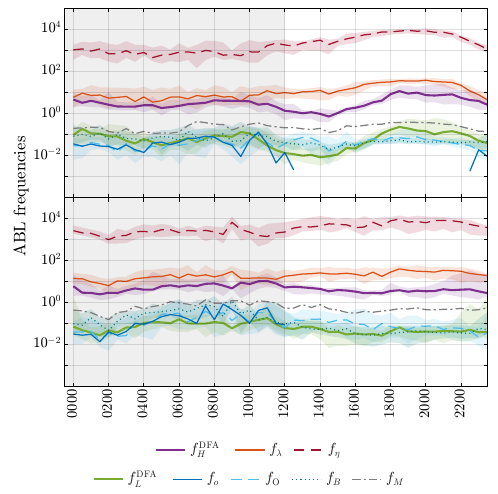}
\caption{Comparison of DFA-derived crossover frequencies with characteristic turbulence scales under stable (shaded intervals) and unstable conditions, at GMT (upper panel) and TMT (lower panel). The high-frequency crossover $f_H^{\mathrm{DFA}}$ (heavy purple) is compared with $f_\lambda$ and $f_\eta$, and the low-frequency crossover $f_L^{\mathrm{DFA}}$ (heavy green) with $f_o$, $f_\mathrm{O}$, $f_B$ and $f_M$.}
      \label{fig:frequencies_scales}
\end{figure}

Low-frequency correspondence between the DFA crossovers and the characteristic frequencies of Sec.~\ref{sec:MOST} depends on the stability regime (Fig.~\ref{fig:frequencies_scales}). Under unstable conditions ($\zeta<0$) the crossover $f_L^\text{DFA}$ aligns with both $f_\text{O}=U/|L|$ and $f_B =U/L_B$, from equations~\eqref{eq:Obukhov_length} and~\eqref{eq:bo_es}, while under stable stratification the low DFA frequencies correspond more closely with the Ozmidov frequency $f_o$, equation~\eqref{eq:oz_es}, than with the Bolgiano frequency. The master-length frequency $f_M = U/L_M$ stays above $f_L^\text{DFA}$ at all hours, on a course of its own.  The buoyancy frequency lies below the other low-frequency scales, at $f_b = (1.67\pm 0.51)\times10^{-2}$~Hz at GMT and $(2.39\pm 0.67)\times10^{-2}$~Hz at TMT, and is omitted from the figure.

At high frequencies the crossover $f_H^\text{DFA}$ corresponds better with the Taylor microscale frequency $f_\lambda$, equation~\eqref{eq:tay_es}, than with the dissipation scale frequency $f_\eta$. The agreement is most pronounced under stable stratification, with deviations reaching up to one order of magnitude during unstable conditions.
\begin{figure*}
   \centering
   \includegraphics[width=\textwidth]{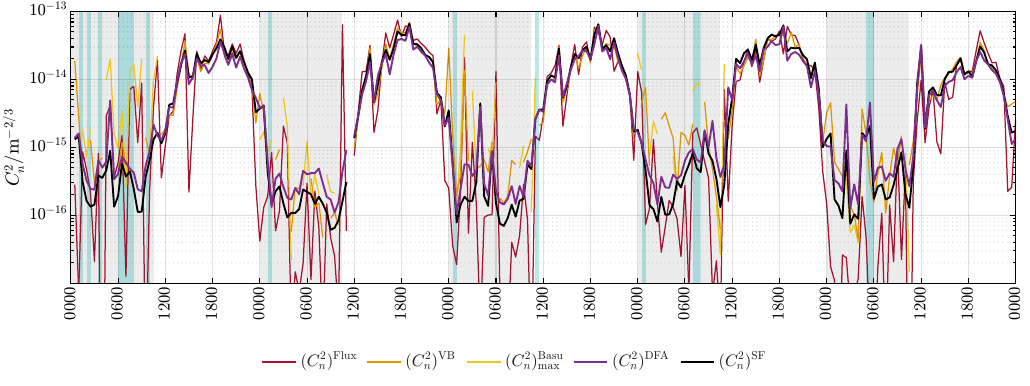}
\caption{Comparison of the five $C_n^2$ estimators of Sec.~\ref{sec:Cn2}, equations~\eqref{eq:strucfun} and~\eqref{eq:genstrucfun}--\eqref{Eq:Cn2_Sukanta}, over five days of the GMT campaign, from December 1st to 5th, 2022. Shaded intervals indicate stable stratification; cyan bands mark records isotropic in the DFA sense.}
   \label{fig:Cn2_Pier}%
\end{figure*}

Figure~\ref{fig:Cn2_Pier} compares the five estimators of Sec.~\ref{sec:Cn2}. Under unstable stratification conditions, when temperature Hurst exponents approach the K41 regime and exhibit coupling with velocity field fluctuations, all $C_n^2$ estimation methods demonstrate excellent agreement. During stable stratification periods, however, methodological discrepancies emerge. The flux-based MOST approach systematically underestimates $C_n^2$ relative to structure function methods by up to two orders of magnitude, while variance-based parameterizations show improved consistency with physical scaling arguments yet still exhibit substantial scatter without systematic bias patterns.

\section{Discussion}
\label{sec:discussion}

The DFA and ARIIS exponents agree in temporal evolution and magnitude, yet correlate only moderately, with the Pearson coefficients of Table~\ref{table:TABLA1} between $0.46$ and $0.74$. Under controlled laboratory conditions, where stationarity holds, DFA and Fourier-based estimates of the generalized Hurst exponent agree at $R^2 > 0.95$ \citep{paper:gonzalez2026}; the gap observed here therefore traces to the atmospheric non-stationarity~\citep{paper:chen2002} that motivates our use of DFA over spectral averaging.

Beyond the candidate inertial-range exponent $H^{(2)}$, DFA identifies the crossover scales $S_{\times}^{(1)}$ and $S_{\times}^{(2)}$ that separate the scaling regimes, information absent in the ARIIS spectral output within its fixed isotropy window. The two criteria explain why the lower bounds of Fig.~\ref{fig:temp_inertial} differ by an order of magnitude. The ARIIS band starts at $U/z$, with campaign medians of $0.12$~Hz at GMT and $0.23$~Hz at TMT, and the isotropy test raises it further, by factors of about three and one and a half; the DFA crossover lies three to four times below $U/z$, where eddies larger than the mast height keep $H\simeq 1/3$ after losing isotropy. At GMT the 30~Hz sampling further limits the high-frequency crossover.

The generalized Hurst exponents of the two methods differ by a systematic offset, DFA exceeding ARIIS by $0.04$ to $0.07$ under unstable stratification and up to about $0.10$ for temperature under stable stratification. This matters in the atmospheric boundary layer, where non-stationarity and stratification-dependent dynamics produce multiple coexisting regimes; spectral averaging smooths these transitions, whereas DFA resolves them directly from the fluctuation function, as Fig.~\ref{fig:clock_exponentes} shows for the stratified flows at both sites. Operating in the time domain, it requires no spectral model and no prior stationarity check \citep{paper:peng1994, paper:kantelhardt2001, paper:kiyono2015}. ARIIS thus provides a spectral reference for the candidate inertial subrange, while DFA maps its physical extent.
 
The departure of the temperature exponent from the K41 value is stronger at TMT, where the sonic sat at $28$~m, than at GMT at $37$~m, with $H_T^{(2)}$ falling below $0.1$ at the lower mast while the velocity exponents hold near $1/3$ at both. Two heights at two sites cannot separate elevation from terrain, yet the direction matches the height dependence \citet{paper:lombardi2010} report at Paranal, where the fraction of temperature spectra with exponents below the Kolmogorov value falls from $79$ to $62$ per cent between $6$ and $30$~m. Proximity to the surface strengthens the buoyancy-driven departure.

A well-established consequence of assuming passive fields is that their structure functions follow the same power law as the velocity field. The decoupling of Fig.~\ref{fig:trend_exponents} therefore suggests that the observed deviations may be attributed to the active behavior of the temperature field (i.e., non-passive), which directly contributes to low-altitude turbulence within the atmospheric boundary layer (ABL). ARIIS recovers the same decoupling, so it survives a change of method and reflects the flow rather than the analysis. The decoupling is also what makes K41 conditions rare: of the $10.5$ and $12.4$ per cent of records isotropic in the DFA sense at GMT and TMT, mostly nocturnal, requiring also $H_T = 1/3 \pm 15\%$ leaves only 6--7 per cent.

The nocturnal values of $H_T^{(2)}$ approaching $\sim 0.2$ or lower align remarkably well with the predictions of \textit{Bolgiano-Obukhov} (BO) scaling theory, which predicts $k^{-7/5}$ spectra in the buoyancy subrange~\citep{paper:bolgiano1959}. The velocity field carries no corresponding signature, and the discrepancy reflects the transition from three-dimensional to quasi-two-dimensional turbulent structures under stable stratification~\citep{paper:basu2019}, where buoyancy-induced vertical density gradients suppress the vertical inertial subrange while potentially preserving horizontal isotropy through \textit{re-entrant} K41 scaling~\citep{paper:boffetta2011}. Consequently, the low-frequency exponent $H_T^{(3)}$ exhibits either persistent BO dynamics or re-entrant K41 behavior at larger scales, depending on the relative strength of buoyancy and mechanical forcing mechanisms. A similar interpretation applies to the shallow TMT slopes, since according to \cite{paper:cheng2020},  spectral exponents approaching $-1.2$ are expected at the transition region between the buoyancy subrange and the isotropic inertial subrange occurring between the buoyancy frequency $f_b$ and the Ozmidov frequency $f_o$.

The nighttime intervals carry flux Richardson numbers $R_f = (g/\overline{\theta})\overline{w'\theta'}/[\overline{u'w'}(\partial u/\partial z)]$ below $0.3$~\citep{book:rstull1988}, indicating subcritical mixing efficiency: shear outweighs buoyancy in turbulent kinetic energy production.  This balance accounts for the absence of the  $k^{-11/5}$ velocity scaling predicted by BO theory, even when temperature spectra show clear BO behaviour. \citet{paper:bolgiano1959} notes that temperature spectra depend on the conversion of turbulent kinetic energy to potential energy, making them more sensitive to buoyancy than velocity in flows with low mixing efficiency---a pattern \citet{paper:chandorrier2024} document for katabatic flows and that \citet{paper:huang2024} report for marine surface layers under weak stable stratification. Active scalar behaviour dominates under stable stratification ($\zeta > 0$), the prevailing condition during astronomical observations, contradicting the passive scalar assumption in conventional optical turbulence characterisation.

The alignment of $f_L^\text{DFA}$ with $f_\text{O}$ and $f_B$ under convective conditions ($\zeta < 0$), reported in Fig.~\ref{fig:frequencies_scales}, indicates that turbulent regime transitions are governed by the interplay between surface-layer dynamics (characterized by $L$) and buoyancy-dominated energy transfer mechanisms. The closer correspondence with the Ozmidov frequency under stable stratification shows that stratification fundamentally alters the cascade dynamics, with the Ozmidov scale providing the critical boundary beyond which three-dimensional isotropy breaks down and anisotropic, quasi-two-dimensional dynamics emerge~\citep{paper:waite2011}. At the high-frequency end, the match with recognised turbulence length scales validates DFA crossover detection as a regime-boundary identifier rather than an empirical fit.

Even under anisotropic conditions meeting our DFA criteria, the tendency toward K41 scaling enables structure constant estimation through structure function approaches $(C_n^2)^\text{SF}$ or $(C_n^2)^\text{DFA}$ combined with Taylor's frozen turbulence hypothesis~\citep{paper:ortiz-suslow2019,paper:lin1953}. Under stable stratification the estimators separate systematically rather than randomly~(Fig.~\ref{fig:Cn2_Pier}), with classical structure function methods deviating by up to two orders of magnitude from flux-based parameterizations, primarily through the breakdown of statistical isotropy and the emergence of active temperature dynamics. The spread measures the sensitivity of the estimators to the regime; telling which one is accurate requires an independent reference. Terrain conditions this separation: over flat ground at $2.28$~m, \citet{paper:muschinski2025} find the methods diverging only at the morning neutral transition, whereas ultrasonic measurements over mountain terrain  at the Sayan Solar Observatory show the structure constant overestimated by more than a factor of two when the Kolmogorov exponent is imposed~\citep{paper:nosov2009}.

Several limitations bound the scope of these results. The analysis rests on single-point measurements at one height per site ($28$~m at TMT, $37$~m at GMT), so the conclusions apply to the surface layer sampled by those masts and not to the full ground layer relevant to adaptive optics. Neither campaign documented anemometer tilt, and the assumption of negligible tilt propagates into the fluxes and hence into $u_*$, $\zeta$, and the flux-based $C_n^2$. The $30$~Hz sampling of the GMT record limits the detection of $S_{\times}^{(1)}$, as Fig.~\ref{fig:temp_inertial} shows, so the high-frequency bounds of the two campaigns are not strictly comparable. The corresponding short-window effects also affect $H^{(1)}$, which lies at scales approaching the dissipative range where self-similar scaling is no longer expected. Although short-scale corrections have been proposed \citep{paper:kantelhardt2001, paper:carpena2021}, leaving them unexplored, we exclude $H^{(1)}$ from the physical interpretation. The generalized constant $(C_n^2)^\text{DFA}$ retains units of m$^{-2H}$ and is therefore a scaling amplitude rather than a physical structure constant; its values are reported for comparison only. Finally, no independent in-situ reference was available at either site, so the intercomparison of Fig.~\ref{fig:Cn2_Pier} establishes the spread among estimators without identifying which lies closest to the temperature statistics. Fine-wire or fibre thermometry, capable of resolving the dissipation range directly, would provide such a reference; optical monitors would not, since they return path-integrated quantities under an assumed spectrum.

\section{Conclusions}
\label{sec:conclusions}

This work set out to determine how often genuine Kolmogorov conditions occur in the ground layer at the Chilean sites of the next generation of extremely large telescopes, and what is lost when they are assumed. Analysis of high-frequency sonic anemometer data from the TMT and GMT campaigns establishes that the three wind components and the temperature simultaneously obey K41 scaling in only 6--7 per cent of observations, when taking a $\pm15$ per cent interval around the K41 exponent. This is too seldom to support universal Kolmogorov statistics in modelling the ABL for optical profiling or AO corrections, although both depend on the whole atmospheric column. DFA, previously validated on controlled optical experiments, transfers to \textit{in situ} anemometry and returns an occurrence rate where none existed, establishing it as a route to optical turbulence characterisation at extremely large telescope sites.

DFA crossover frequencies correspond to characteristic turbulence scales, the low-frequency crossover aligning with the Obukhov-length and Bolgiano frequencies under unstable conditions and tracking the Ozmidov frequency under stable stratification, while the high-frequency crossover coincides with the Taylor microscale transition. The bounds of the inertial subrange are therefore fixed on physical rather than numerical grounds, which validates DFA as a diagnostic that complements and extends spectral approaches such as ARIIS in non-stationary boundary layer flows.

Stratified conditions decouple velocity and temperature scaling exponents (Fig.~\ref{fig:trend_exponents}). Velocity components stay near Kolmogorov through parts of the diurnal cycle, while temperature exhibits Bolgiano-Obukhov scaling ($H \approx 0.2$, $k^{-7/5}$ spectra) during roughly 60 per cent of nighttime intervals (Fig.~\ref{fig:clock_exponentes}). Therefore, active scalar behaviour dominates under stable stratification, the prevailing condition during astronomical observations, and the passive-scalar assumption on which conventional $C_n^2$ estimation rests fails precisely when it matters most. Classical structure function methods and flux-based parameterizations then separate by up to two orders of magnitude, a spread that no choice among them resolves without knowing which regime the flow is in. 

The rarity of K41 conditions at world-class observatory sites indicates that routine K41-based profiling mischaracterises surface turbulence. For an optical profiler, a seeing budget built on either approach to $C_n^2$ alone  inherits an uncertainty of comparable magnitude under the conditions in which the observations are made. What future profiling instruments at the next-generation observatories require is a means of recognising when the classical assumptions fail, and DFA supplies it, resolving the regime from the record itself rather than presuming one.

\section*{Acknowledgements}

We thank the anonymous reviewer, whose report improved the structure of this paper. We thank Warren Skidmore from TMT International Observatory for providing the site testing campaign sonic anemometer data. We are also especially grateful to Rodolphe Conan and Francisco Meza from GMTO Organization for coordinating and running the campaign at Las Campanas Observatory. This work was financed by Agencia Nacional de Investigaci\'{o}n y Desarrollo (ANID), Chile (FONDECYT 1211848; ANILLO ATE250061; FONDEF ID26I10339); Pontificia Universidad Cat\'{o}lica de Valpara\'{i}so (PUCV), Chile.  The authors declare no conflict of interest.

\section*{Data Availability}

The GMT sonic anemometer data underlying this article were collected during a campaign coordinated by GMTO Organization at Las Campanas Observatory; requests for access should be directed to GMTO. The TMT sonic anemometer data were provided by TMT International Observatory and are subject to their data sharing policies; requests should be directed to TMT International Observatory.

\bibliographystyle{mnras}
\bibliography{gsm}

\bsp
\label{lastpage}
\end{document}